\documentclass[sigconf,screen,authorversion=true]{acmart}

\usepackage{balance}
\usepackage{amsmath,amssymb,amsfonts}
\usepackage{algorithmic}
\usepackage{graphicx}
\usepackage{textcomp}
\usepackage{xcolor, colortbl}
\usepackage{paralist}
\usepackage{enumitem}
\usepackage{soul}
\usepackage{tcolorbox}
\usepackage{booktabs, multirow, pgfplots}
\usepackage{framed}
\usepackage{xspace}

\newcommand{\ie}{i.e.,\xspace}
\newcommand{\eg}{e.g.,\xspace}

\newcommand{\etal}{et al.\xspace}

\newcommand{\mozilla}{\textsc{Mozilla}\xspace}

\newcommand{\nresponses}{121\xspace}
\newcommand{\nprofessionals}{106\xspace}
\newcommand{\nmozilladevs}{21\xspace}
\newcommand{\nflakytests}{200\xspace}
\newcommand{\nnewcauses}{four\xspace}

\newcommand{\rqOne}{How do professional developers categorize flaky tests, in terms of nature, origin, and fixing effort? How do they fix them?\xspace}
\newcommand{\rqTwo}{How prominent is test flakiness and how problematic is it as perceived by developers?\xspace}
\newcommand{\rqThree}{What are the main challenges that developers face when dealing with flaky tests?}

\definecolor{gray50}{gray}{.5}
\definecolor{gray40}{gray}{.6}
\definecolor{gray30}{gray}{.7}
\definecolor{gray20}{gray}{.8}
\definecolor{gray10}{gray}{.9}
\definecolor{gray05}{gray}{.95}

\newlength\Linewidth
\def\findlength{\setlength\Linewidth\linewidth
	\addtolength\Linewidth{-4\fboxrule}
	\addtolength\Linewidth{-3\fboxsep}
}
\newenvironment{examplebox}{\par\begingroup
	\setlength{\fboxsep}{5pt}\findlength
	\setbox0=\vbox\bgroup\noindent
	\hsize=0.95\linewidth
	\begin{minipage}{0.95\linewidth}\normalsize}
	{\end{minipage}\egroup
	\textcolor{gray20}{\fboxsep1.5pt\fbox
		{\fboxsep5pt\colorbox{gray05}{\normalcolor\box0}}}
	\endgroup\par\noindent
	\normalcolor\ignorespacesafterend}

\newcommand{\rb}[1]{
	
	\vspace{0.3cm}
	\begin{tcolorbox}[colback=gray!05,
		colframe=black,
		width=\columnwidth,
		arc=3mm, auto outer arc,
		boxrule=0.5pt,
		]
		#1
	\end{tcolorbox}
}

\newcounter{Finding}
\stepcounter{Finding}

\newcommand{\roundedbox}[1]{
	\rb{
		\noindent
		\emph{\textbf{Finding \theFinding}. #1}
	}
	\stepcounter{Finding}
}

\usepackage{listings}
\usepackage{color}
\usepackage{xcolor}

\definecolor{dkgreen}{rgb}{0,0.6,0}
\definecolor{gray}{rgb}{0.5,0.5,0.5}
\definecolor{mauve}{rgb}{0.58,0,0.82}

\lstdefinelanguage{diff}{
	basicstyle=\scriptsize,
	morecomment=[f][\color{diffstart}]{@@},
	morecomment=[f][\color{diffincl}]{+\ },
	morecomment=[f][\color{diffrem}]{-\ },
}

\setcopyright{acmlicensed}
\copyrightyear{2019} 
\acmYear{2019} 
\acmConference[ESEC/FSE '19]{Proceedings of the 27th ACM Joint European Software Engineering Conference and Symposium on the Foundations of Software Engineering}{August 26--30, 2019}{Tallinn, Estonia}
\acmBooktitle{Proceedings of the 27th ACM Joint European Software Engineering Conference and Symposium on the Foundations of Software Engineering (ESEC/FSE '19), August 26--30, 2019, Tallinn, Estonia}
\acmPrice{15.00}
\acmDOI{10.1145/3338906.3338945}
\acmISBN{978-1-4503-5572-8/19/08}


\begin{document}

\title{Understanding Flaky Tests: The Developer's Perspective}

\author{Moritz Eck}
\orcid{}
\affiliation{%
  \institution{University of Zurich}
  \city{Zurich}
  \country{Switzerland} 
}
\email{moritz.eck@uzh.ch}

\author{Fabio Palomba}
\orcid{0000-0001-9337-5116}
\affiliation{%
  \institution{University of Zurich}
  \city{Zurich}
  \country{Switzerland}
}
\email{palomba@ifi.uzh.ch}

\author{Marco Castelluccio}
\orcid{}
\affiliation{%
  \institution{Mozilla Software Foundation}
  \city{London}
  \country{United Kingdom} 
}
\email{mcastellucio@mozilla.org}

\author{Alberto Bacchelli}
\orcid{0000-0003-0193-6823}
\affiliation{%
  \institution{University of Zurich}
  \city{Zurich}
  \country{Switzerland} 
}
\email{bacchelli@ifi.uzh.ch}

\begin{abstract} 
 Flaky tests are software tests that exhibit a seemingly random outcome (pass or fail) despite exercising unchanged code.
 In this work, we examine the perceptions of software developers about the nature, relevance, and challenges of flaky tests.
 
 We asked \nmozilladevs professional developers to classify \nflakytests flaky tests they previously fixed, in terms of the nature and the origin of the flakiness, as well as of the fixing effort. We also examined developers' fixing strategies.
 Subsequently, we conducted an online survey with \nresponses developers with a median industrial programming experience of five years.
 Our research shows that:
 The flakiness is due to several different causes, \nnewcauses of which have never been reported before, despite being the most costly to fix;
 flakiness is perceived as significant by the vast majority of developers, regardless of their team's size and project's domain, and it can have effects on resource allocation, scheduling, and the perceived reliability of the test suite;
 and the
 challenges developers report to face regard mostly the reproduction of the flaky behavior and the identification of the cause for the flakiness.
 Data and materials [\url{https://doi.org/10.5281/zenodo.3265785}].
\end{abstract}
	
	
\maketitle

\section{Introduction}
\label{sec:introduction}

Software tests are \emph{flaky} when they exhibit a seemingly random outcome despite exercising code that has not been changed.

Even though previous work has proposed techniques to automatically fix some types of flaky tests~\cite{Bell:2014,bell2018deflaker,Farchi:2003,Jin:2011,Zhang:2014:ERT}, our scientific knowledge about flaky tests is still very limited.

Luo \etal~\cite{Luo:fse2014} presented the earliest and most significant work advancing our empirical knowledge on flaky tests. The researchers inspected 201 commits that likely fixed flaky tests from 51 Apache projects. They, thus, derived a taxonomy of the most common root causes (their \emph{nature}, henceforth) of flaky tests, as well as identified strategies to manifest and fix certain types of flakiness~\cite{Luo:fse2014}.

The work we present in this paper continues on this line of research about understanding flaky tests. Here we present \emph{the developer's perspective}: Our goal is to investigate developers' perception on the causes and effort in fixing flaky tests, the significance of the problem, as well as what they deem to be the most important challenges. An improved comprehension of these aspects is key to the definition of solutions and research lines to better help practitioners in diagnosing and fixing test flakiness.

To achieve our goal, \begin{inparaenum}[(1)]
	\item we ask \nmozilladevs professional \mozilla developers to classify \nflakytests real-world flaky tests they had previously fixed, in terms of the nature of the flakiness, the origin of the flakiness (test or production code), and the fixing efforts. We complement this analysis with information about the fixing strategy, which we collect from the source code repository. Focusing on a single ecosystem gives us the opportunity to detect the less frequent cases that would not likely appear by panning wide.
	Subsequently,
	\item we conduct an online survey that was answered by \nresponses developers (\nprofessionals professionals and 15 academics). Since these developers are from different projects and backgrounds, their answers give us the opportunity to learn from a variety of cases.
\end{inparaenum}

The categorization by the \mozilla developers uncovered \nnewcauses previously unreported causes of flakiness, which are also deemed as those requiring the most effort to fix. Most surveyed developers (79\%) consider flaky tests a moderate to serious problem, regardless of their team's size and project's domain; and 40\% deal with flaky tests at least weekly. In terms of problems, flaky tests are reported to have serious consequences on the scheduling, allocation, and reliability of the testing process. Finally, reproducing the flaky behavior and classifying its cause are perceived as the major challenges.

We publicly release the full dataset concerning our contributions, as well as the data and materials for the futher analyses~\cite{eck:appendix}.

\section{Goals and Method's Overview}
\label{sec:rqs}

The goal of this study is to capture's the developer's perspective on flaky tests with the purpose of: \begin{inparaenum}[(1)]
	\item understanding the nature of test flakiness, investigating why and where flaky tests arise, as well as the fixing effort and strategies,
	\item gathering the relevance of and problems caused by flaky tests in practice,
	and
	\item collecting the challenges faced by developers when handling flaky tests.
\end{inparaenum}

\subsection{Research Questions}
We start our research by asking professional developers to classify (in terms of nature, origin, and fixing effort) flaky tests they previously fixed. Differently from Luo \etal \cite{Luo:fse2014}, we do not classify the underlying causes of flakiness ourselves, rather the original developers do and add information on the origin and fixing effort. Moreover, we complement the developers' analysis with information on the fixing strategies. Our goal is to triangulate the taxonomy proposed by Luo \etal \cite{Luo:fse2014} from a different data source and with a different methodology.
Hence, our first research question:

\begin{center}	
	\begin{examplebox}
		\textbf{RQ$_1$.} \emph{\rqOne}
	\end{examplebox}	 
\end{center}

Subsequently, we turn to a broader audience of respondents. We conduct a survey targeting software developers on their experience with flaky tests.
We investigate the frequency of flaky tests in practice and how problematic flaky tests are from the developer's perspective. 
We ask:

\begin{center}	
	\begin{examplebox}
		\textbf{RQ$_2$.} \emph{\rqTwo}
	\end{examplebox}	 
\end{center}

Finally, we investigate the challenges that developers perceive as most critical when they have to deal with flaky tests. We ask:

\begin{center}	
	\begin{examplebox}
		\textbf{RQ$_3$.} \emph{\rqThree}
	\end{examplebox}	 
\end{center}

\subsection{Overview Of The Research Method}
Our study features a mixed-methods approach \cite{johnson2004mixed} where quantitative and qualitative research are run in parallel with the goal of converging toward an empirical understanding of test code flakiness. We design and conduct a study to obtain data from two main sources: A novel annotated dataset (\textbf{\textsc{ds}}) of flaky tests (with \nflakytests instances annotated by \nmozilladevs professional developers) and the responses to an online survey (\textbf{\textsc{os}}) (with \nresponses valid responses from developers).

\subsection{Developers' Analysis Of Flaky Tests}

In the following, we describe the methodological steps conducted to address our first research question.\smallskip

\textbf{Subject Flaky Tests.}
We first need to collect a dataset of flaky tests, which we then present to developers for their analysis.
To this aim, we mine the \textsc{Bugzilla} issue tracker of \textsc{Mozilla} (\ie a large free software organization counting more than 40,000 active contributors spread around the world and developing a variety of software, ranging from layout engines to operating systems). \textsc{Mozilla} has a database of flaky tests, verified and fixed by developers. According to such a database, the organization has 100 to 150 new flaky tests being detected every week: this makes it particularly suitable for our study because developers are very frequently in charge of diagnosing and fixing flaky tests, being therefore able to provide authoritative information on the nature of test flakiness, its origin, and the fixing effort.

We extract the data available after a flaky test has been reported. Thus, we gather the log files produced by the continuous integration system and the source code associated with the fixed flaky test (if available). 
In a first step, we identify all (869) fixed flaky tests the \mozilla's bug tracking system tagged \texttt{resolved} and \texttt{fixed} between May 2017 and April 2018. We consider this time window since the company only stores the log files for twelve months after their creation. 
From the set of retrieved flaky tests, we remove all instances for which no patch has been applied or other type of attachment is available (\eg some flaky tests stop occurring after a while and are closed as fixed), as they were never fixed or were false alarms.
This step filters out 295 tests; resulting in a set of 574 flaky tests, which we use in the next steps.

\smallskip
\textbf{Recruiting \mozilla Developers.}
The remaining 574 actual flaky tests are taken into further consideration and sorted according to the developer who is listed as the assignee or patch creator. From this list, to ensure we select people with a concrete knowledge on flaky tests, we exclude 177 programmers who have only fixed less than five  flaky tests. We end up with a list of 29 developers for a total of 304 fixed flaky tests. We contact these programmers through email and ask them to analyze the flaky tests as detailed in the following subsection.

\smallskip
\textbf{Developers' Analysis.}
We create a spreadsheet for each developer that contains two columns: the first reports the link to the bug report related to each flaky test fixed by a certain developer; the second is used by the programmer to classify the nature of each flaky test in the list. The participating developers are allowed to use the taxonomy defined by Luo \etal~\cite{Luo:fse2014} as a starting point for the classification process, as it (1) eases their task and (2) standardizes their answers---this is commonly done in both confirmatory and exploratory surveys \cite{flanigan2008conducting}. 
However, they are also allowed to create additional categories if none of the existing ones fit; at the same time, they can indicate more than one type to specify the reasons behind the flakiness of a test (it may be flaky because of both a race condition and a network problem).
We also ask the developers to specify the development effort they spent to fix the flaky tests, based on a Likert scale \cite{Oppenheim:1992} between 1 (very low) to 5 (very high), as well as the origin (test or production code) of the flakiness.
Once each developers completes the analysis, they send back their spreadsheets with the annotated classification. 
We received a total of 248 responses from \nmozilladevs professional \mozilla developers. These developers have 7.5 years (median) of experience with \mozilla, 958 bug reports (median) assigned to, and---over the last year---14 flaky tests fixed (median). Each developer provided us with 8.5 (median) answers (min = 4, max = 48). Of those responses, we excluded 48: (i) 24 cases turned into permanent failures, \ie they were labeled as intermittent by a \mozilla sheriff~\cite{mozillaSheriff}, but the fixing developer found them to be permanent failures; (ii) in the other 24 cases, the type of flakiness was unknown even to the developers, meaning that the fixing developers worked around the flakiness rewriting the whole test and code-under-test, but they could not ascertain the cause of the flakiness. By removing these cases, we intend to rely on precise information, only. Nevertheless, the excluded cases are available to our online appendix~\cite{eck:appendix}.
\smallskip

\textbf{Analysis Of The Developers' Responses.}
We use the \nflakytests classified flaky tests to define a taxonomy on the nature of flaky tests. In 72 cases (31\%) the assigned categories do not correspond to those available in the taxonomy by Luo \etal~\cite{Luo:fse2014} and, therefore, they are new and need to be defined. To this aim, we conduct an \emph{iterative content analysis} \cite{white2006content} to assign a common name to these new categories. The process involves three software engineering researchers, all authors of this paper, (one graduate student, one research associate, and one faculty member) and consists of the two iterative sessions:

\begin{description}[leftmargin=0.3cm]
	\item[Iteration 1:] The first author of this paper goes over the classifications made by developers. If the categories assigned belong to the taxonomy by Luo \etal~\cite{Luo:fse2014}, he leaves them as they are; if not, he assigns a temporary label to them. As an output, this step provides a draft categorization of the types of flakiness. 
	
	\smallskip
	\item[Iteration 2:] The three researchers open a discussion on the draft taxonomy, with the aim of reaching a consensus on the assigned categories. Afterwards, the first author re-categorizes flaky tests according to the decisions taken during the discussion. Finally, the second author of the paper double-checks the classifications to reduce the risk of errors (he found no wrong classifications, thus reaching a total agreement).
	
\end{description}

The resulting extended taxonomy is then used to address \textbf{RQ$_1$}. Besides reporting the list of such categories, we also characterize them with additional information on \begin{inparaenum}[(i)]
	\item their frequency, computed on the basis of the distribution of the categories within the dataset, 
	and 
	\item the fixing strategies.
\end{inparaenum}
As for the latter point, to characterize the fixing strategies adopted by developers when dealing with flaky tests, two authors of this paper conduct a new iterative content analysis that---similarly to the previous one---consists of two iterations. In the first iteration, the first author analyzes the patches associated to the resolution of the considered flaky tests to label them with a name and a short description of the fixing action performed by developers. In the second iteration, the two authors open a discussion on the initial labels assigned to reach a consensus. Then, the first author applies the changes according to the discussion while the second re-checks the final classifications to reduce threats in the interpretation of the results.

\subsection{Survey: Research Method}\label{sec:survey_method}

To answer \textbf{RQ$_2$} and \textbf{RQ$_3$}, we design and deploy an online survey, and analyze the collected answers.

\smallskip
\textbf{Overall Survey Design.} Following the guidelines provided by Flanigan \etal~\cite{flanigan2008conducting}, we limit common issues possibly arising in survey studies and affecting the response rate: we keep the survey short, respecting the anonymity of participants and preventing our influence in the answers. We create an anonymous online survey using a professional tool (\ie \textsc{Surveygizmo}~\cite{surveygizmo}).

The survey first describes to the respondents the concept of flaky tests, following the definition accepted by the research community: \emph{``[flaky tests are] intermittent tests that sometimes pass, sometimes fail, even though there are no changes in the code they test''}. Then, the survey 
asks for demographic information about the participants, including programming/testing experience as well as company and team size/domain. Subsequently, the survey contains other two main sections to collect data about the relevance of flakiness (\textbf{RQ$_2$}) and its challenges (\textbf{RQ$_3$}).\smallskip

\textbf{\textbf{RQ$_2$} -- Collecting Information About Relevance.} We gather data on \begin{inparaenum}[(i)]
	\item how many times flaky tests occur,
	\item how problematic they are,
	and
	\item what are the top 3 to 5 problems caused by test flakiness.
\end{inparaenum}
Participants can answer the first two questions by using a 5-point Likert scale indicating the extent to which the problem exists and how much it is serious, while they are free to write down the problems (if any) in a text box.\smallskip

\textbf{\textbf{RQ$_3$} -- Collecting Information About Challenges.}

We first conduct a multivocal literature review (MLR)~\cite{garousi2016need}. This allows us to let emerge eight pieces of information that may be linked to challenges in the process of fixing flaky tests.
Afterwards, we ask the survey respondents to rate how important is each of those information pieces from `Not at all important' to `Extremely important' (on a 4-point Likert scale) and the difficulty in obtaining it from `Very easy' to `Very difficult' (on a 5-point Likert scale). We also let the respondents indicate any information needs and/or challenges they face that are not included in the list (respondents indicated a total of another eleven needs/challenges).

For our MLR, we analyze (as described in the following protocol) both previously published papers on the topic (\ie white literature) and online sources (\eg blog posts, websites, and documents written by practitioners) related to flaky tests (\ie gray literature).

\begin{itemize}[leftmargin=0.3cm]
	\item \emph{Identifying relevant white literature:} We perform a keyword search on \textsc{Google Scholar}, \textsc{IEEE Explore}, and \textsc{ACM Digital Library} to create an initial selection of articles. As keywords we use: `flaky tests', `flaky test resolution practices', `intermittent failures'. This step results in an initial set of 15 papers.
	We also perform forward and backward snowballing \cite{wohlin2014guidelines}, inspecting sources referred or that refer to those belonging to the initial set of primary studies. 
	At the same time, we consider the proceedings of all the relevant, top-tier conference venues in Software Engineering such as ICSE \cite{icse}, ISSTA \cite{issta}, ICST \cite{icst}, ICSME \cite{icsme}, ESEC/FSE \cite{fse} and journals such as IEEE TSE \cite{tse}, ACM TOSEM \cite{tosem}, and Springer's EMSE \cite{emse}. This step results in the inclusion of additional four papers.
	Once having this set of resources, we filter out the papers that do not report information on what are the practices used by developers to diagnose and/or fixing flaky tests: This is achieved by reading study setting, methodology, and conclusions made in all the considered papers. The filtering process is jointly discussed among all the authors of the paper and, in the end, we agree on a list of three papers. 
	
	\smallskip
	\item \emph{Identifying relevant gray literature:} We conduct the gray literature survey following the guidelines by Garousi \etal \cite{garousi2016need}. We query \textsc{Google} by using `flaky tests', `flaky test resolution practices', `intermittent failures' as keywords. We perform our search on the results appearing in all the relevant \textsc{Google} pages until saturation is reached, \ie until the search results contain references to blogs, documents, or websites reporting information on flaky tests. We take great care to ensure that considered sources originate from reputable companies, organizations, and authors. We also take steps from preventing our personal search bias from playing a role in the search results, hence we search on \textsc{Google} using the incognito mode. Overall, we find 56 relevant sources to further consider. Whenever possible, we also follow the web links contained in the initial set of sources as they might refer to other documents. This step results in the addition of three sources. The high number of relevant sources identified in the gray literature, as opposed to that of the white one, hints at the fact that the problem of understanding flaky tests has been only partially addressed by the research community. This situation calls for further empirical studies to create a stronger scientific knowledge around test flakiness.
	
	\smallskip
	\item \emph{Data Extraction:} We summarize each of the 62 identified relevant sources and create an enumeration of the aspects that are mentioned and may impact on the flaky tests fixing process. Then, we conduct another iterative content analysis sessions to group those aspects into the higher level themes that are related to the important information developers need to have when diagnosing and fixing test flakiness. This is a two-step process similar to the one conducted in the context of \textbf{RQ$_1$}: the second author of this paper performs a first iteration and then the draft themes are refined with the help of the last author. This iterative process results in eight relevant information types, which we list in the survey as previously mentioned.
	
\end{itemize}

\textbf{Analysis Of The Open Answers.}
For the open answers in the survey, we conduct a two-stage process comprising Descriptive and Pattern coding analysis to name them \cite{corbin1990grounded}. Specifically, in a first step two authors shortly summarize the topic of each passage from the open answers; then, they identify explanatory codes to create themes that are finally discussed among all the authors. For example, in case of the challenges, with this process we end up with nine additional challenges grouped into two categories that indicate the point in time in which they occur, \ie (1) \emph{Test case design} and (2) \emph{Flaky test fixing}.

\smallskip
\textbf{Attracting Participants.} We advertised the online survey using the personal social network accounts of the authors (\ie \textsc{Facebook}, \textsc{Twitter}, and \textsc{LinkedIn}), through private contacts, and also published on practitioners blogs (\eg \textsc{Reddit}). To stimulate the participation, we allow the participants to indicate a non-profit organization of their choice to which we donate 2 USD. 

\smallskip
\textbf{Respondents.} As a result, we collect 188 answers: We filter out 68 partial answers, thus, we only consider \nresponses full responses. Among these respondents, 75\% have more than 6 years of programming experience (median 10) and two years of industrial programming as well as testing experience (median 5). Moreover, 45\% respondents come from large companies having more than 250 employees. Our sample is composed of developers who are experienced with (industrial) programming and tests, thus we consider the collected responses as valuable for our research questions.

\section{\textbf{RQ$_1$} -- Flaky Tests: Nature, Origin, and Fixing}
\label{sec:rq1}

The first research question aims at describing the possible types of flakiness, as well as their origin, fixing effort and strategies.

\begin{table*}[!h]
	\centering
	\caption{Taxonomy of the nature of flaky tests with the corresponding fixing strategies, by frequency (N = 234, because developers were allowed to assign more than one nature to each of the \nflakytests flaky tests they analyzed). The `*' and underlining indicates newly reported categories (\ie they were not included in the taxonomy proposed by Luo \etal \cite{Luo:fse2014}).}
	\label{tab:fixingTaxonomy}
	\resizebox{0.85\linewidth}{!}{
		\begin{tabular}{|llp{10cm}r|}
			\hline

			\cellcolor{black}\textcolor{white}{Nature: Concurrency} & \multicolumn{3}{l|}{{Cases: 61 \hspace*{0.25cm}|\hspace*{0.25cm} Avg. Fixing Effort: 4.0 \hspace*{0.25cm}|\hspace*{0.25cm} Origin: \phantom{1}66\% Test Code}}\\\hline
			\rowcolor{gray!50} \multicolumn{4}{|l|}{Fixing Strategies:} \\
			\rowcolor{gray!50} \multicolumn{2}{|l}{Name} & Description & Frequency \\
			\multicolumn{2}{|l}{Wait For Addition} & An \texttt{await} or \texttt{waitFor} promise is added to wait for the required event. & 46\% \\ 
			\rowcolor{gray!20} \multicolumn{2}{|l}{Change Concurrency Guard Conditions}  & Modification of the constraints that may block the applicability of the clause under some conditions. & 26\% \\
			\multicolumn{2}{|l}{Adding Lock} & New locks are introduced to ensure mutual exclusion between threads. & 21\% \\
			\rowcolor{gray!20} \multicolumn{2}{|l}{Make Code Deterministic} & The concurrency is removed. & 5\% \\
			\multicolumn{2}{|l}{Test Replacement} & The flaky test is replaced with a brand new test. & 1\% \\
			\rowcolor{gray!20} \multicolumn{2}{|l}{Disable Test} & The flaky test is disabled and no longer executed. & 1\% \\\hline\hline
			
			\cellcolor{black}\textcolor{white}{Nature: Async Wait} & \multicolumn{3}{l|}{{Cases: 52 \hspace*{0.25cm}|\hspace*{0.25cm} Avg. Fixing Effort: 3.0 \hspace*{0.25cm}|\hspace*{0.25cm} Origin: 100\% Test Code}}\\\hline
			\multicolumn{2}{|l}{Wait For Addition} & An \texttt{await} or \texttt{waitFor} promise is added to wait for the required event. & 86\% \\ 
			\rowcolor{gray!20} \multicolumn{2}{|l}{Reordering Threads Execution} & The source code is reorganized or refactored to make the execution deterministic or less async. & 13\% \\ 
			\multicolumn{2}{|l}{Disable Test} & The flaky test is disabled and no longer executed. & 1\% \\\hline\hline

			\cellcolor{black}\textcolor{white}{Nature: \underline{*Too Restrictive Range}} & \multicolumn{3}{l|}{{Cases: 40 \hspace*{0.25cm}|\hspace*{0.25cm} Avg. Fixing Effort: 1.0 \hspace*{0.25cm}|\hspace*{0.25cm} Origin: 100\% Test Code}}\\\hline
			\multicolumn{2}{|l}{Fix Assertion} & The cause leading to the failure is diagnosed and fixed. & 45\% \\ 
			\rowcolor{gray!20} \multicolumn{2}{|l}{Adjust Fuzzing} & The \texttt{assert} statement is modified so that it provides less extreme values or allow larger difference when comparing expected and actual outcome. & 20\% \\ 
			\multicolumn{2}{|l}{Disable Test} & The flaky test is disabled and no longer executed. & 16\% \\
			\rowcolor{gray!20} \multicolumn{2}{|l}{Missing Code Added} & The code needed to check whether a certain condition is met is added. & 1\% \\\hline\hline

			\cellcolor{black}\textcolor{white}{Nature: Test Order Dependency} & \multicolumn{3}{l|}{{Cases: 22 \hspace*{0.25cm}|\hspace*{0.25cm} Avg. Fixing Effort: 2.0 \hspace*{0.25cm}|\hspace*{0.25cm} Origin: 100\% Test Code}}\\\hline
			\multicolumn{2}{|l}{Remove Dependency} & Changes aimed at (i) modifying the output directory for the test to use a separate directory or (ii) checking instance variables before that the test is accessed and executed. & 100\% \\\hline\hline
			
			\cellcolor{black}\textcolor{white}{Nature: \underline{*Test Case Timeout}} & \multicolumn{3}{l|}{{Cases: 18 \hspace*{0.25cm}|\hspace*{0.25cm} Avg. Fixing Effort: 4.0 \hspace*{0.25cm}|\hspace*{0.25cm} Origin: 100\% Test Code}}\\\hline
			\multicolumn{2}{|l}{Increase Timeout} & The timeout time is increased to avoid the flakiness behavior. & 85\% \\
			\rowcolor{gray!20} \multicolumn{2}{|l}{Skip Non-Initialized Part} & Code added to skip non-initialized parts to make the test run faster and not timeout. & 5\% \\
			\multicolumn{2}{|l}{Split Test} & Split up all tests (run in parallel) into more groups to reduce risk of a timeout. & 5\% \\
			\rowcolor{gray!20} \multicolumn{2}{|l}{Disable Test} & The flaky test is disabled and no longer executed. & 5\% \\\hline\hline
			
			\cellcolor{black}\textcolor{white}{Nature: Resource Leak} & \multicolumn{3}{l|}{{Cases: 14 \hspace*{0.25cm}|\hspace*{0.25cm} Avg. Fixing Effort: 3.0 \hspace*{0.25cm}|\hspace*{0.25cm} Origin: \phantom{1}85\% Test Code}}\\\hline
			\multicolumn{2}{|l}{Destroy Object} & New code is added so that the conflicting object is destroyed before continuing the execution. & 50\% \\
			\rowcolor{gray!20} \multicolumn{2}{|l}{Test Replacement} & The flaky test is replaced with a brand new test. & 36\% \\
			\multicolumn{2}{|l}{Disable Test} & The flaky test is disabled and no longer executed. & 14\% \\\hline\hline
			
			\cellcolor{black}\textcolor{white}{Nature: \underline{*Platform Dependency}} & \multicolumn{3}{l|}{{Cases: 10 \hspace*{0.25cm}|\hspace*{0.25cm} Avg. Fixing Effort: 4.0 \hspace*{0.25cm}|\hspace*{0.25cm} Origin: \phantom{1}90\% Test Code}}\\\hline
			\multicolumn{2}{|l}{Run Additional Tests} & New platform-specific tests are ran instead of the flaky one. & 60\% \\
			\rowcolor{gray!20} \multicolumn{2}{|l}{Correct Directories} & The test suite is modified so that it includes the correct platforms to run the tests on or to fix a directory issue. & 40\% \\\hline\hline	
			
			\cellcolor{black}\textcolor{white}{Nature: Float Precision} & \multicolumn{3}{l|}{{Cases: \phantom{1}6 \hspace*{0.25cm}|\hspace*{0.25cm} Avg. Fixing Effort: 4.0 \hspace*{0.25cm}|\hspace*{0.25cm} Origin: 100\% Test Code}}\\\hline
			\multicolumn{2}{|l}{Subtract Bytecode Offset} & The floating point is corrected by subtracting the bytecode offset so that the precision is preserved. & 100\% \\\hline\hline	
			
			\cellcolor{black}\textcolor{white}{Nature: \underline{*Test Suite Timeout}} & \multicolumn{3}{l|}{{Cases: \phantom{1}4 \hspace*{0.25cm}|\hspace*{0.25cm} Avg. Fixing Effort: 3.5 \hspace*{0.25cm}|\hspace*{0.25cm} Origin: 100\% Test Code}}\\\hline
			\multicolumn{2}{|l}{Skip Non-Initialized Part} & Code added to skip non-initialized parts to make the test run faster and not timeout. & 75\% \\
			\rowcolor{gray!20} \multicolumn{2}{|l}{Split Test Suite} & Split up all tests (run in parallel) into more groups to reduce risk of a timeout. & 25\% \\\hline\hline 
			
			\cellcolor{black}\textcolor{white}{Nature: Time} & \multicolumn{3}{l|}{{Cases: \phantom{1}4 \hspace*{0.25cm}|\hspace*{0.25cm} Avg. Fixing Effort: 1.0 \hspace*{0.25cm}|\hspace*{0.25cm} Origin: 100\% Test Code}}\\\hline
			\multicolumn{2}{|l}{Disable Test} & The flaky test is disabled and no longer executed. & 75\% \\
			\rowcolor{gray!20} \multicolumn{2}{|l}{Preserve Time Precision} & The test case is modified so that the time precision is preserved. & 25\% \\\hline\hline	
			
			\cellcolor{black}\textcolor{white}{Nature: Randomness} & \multicolumn{3}{l|}{{Cases: \phantom{1}3 \hspace*{0.25cm}|\hspace*{0.25cm} Avg. Fixing Effort: 1.0 \hspace*{0.25cm}|\hspace*{0.25cm} Origin: 100\% Test Code}}\\\hline
			\multicolumn{2}{|l}{Replaced Math.random} & The \texttt{Math.random} call is replaced with more reliable random number generator. & 100\% \\\hline
									
		\end{tabular}
	}
\end{table*}
The categories are reported by frequency of nature and, within those, by frequency of fixing strategies. For consistency with previous literature, whenever possible, we followed the taxonomy by Luo \etal \cite{Luo:fse2014}; newly reported categories (\ie natures of flakiness that were not included in the taxonomy proposed by Luo \etal \cite{Luo:fse2014}) are underlined and marked with a `*'.

\smallskip
\begin{description}[leftmargin=0.3cm]

	\item[Concurrency.] A test that is flaky due to synchronization issues originating from multiple threads interacting in an unsafe manner is classified in this category. For example, a race condition occurs due to threads relying on an implicit ordering of the execution leading to a deadlock in certain situations. As observed by Luo \etal~\cite{Luo:fse2014}, this is a high diffused cause of flakiness. In our dataset, it is present in 61 cases (26\%). Moreover, the median effort required to fix this problem---as assigned by developers---is 4 (high), indicating that it is problematic and further justifying the amount of research effort done so far to define automatic techniques able to solve it \cite{Memon:2013,Muslu:2011}. Interestingly, not all flaky tests in this category origin in test code: indeed, the developers report that in 34\% of the cases the fixing process requires the examination of the production code and not of the test. Thus, \emph{test flakiness can be originated by the production code}; besides motivating further research on the origin of flaky tests, this finding indicates that automated approaches aimed at fixing this problem must consider whether the flakiness originates in test or production code. As a final note, there seem to be one key fixing strategy, namely the addition of a \texttt{waitFor} statement, while other operations, like the modification of the concurrency guard condition or the addition of \texttt{lock} statement to avoid other threads running at the same time represent less frequent strategies to remove this flakiness. In 5\% of the cases, developers act by refactoring production code to remove concurrency, thus making the code deterministic. Finally, it is also interesting to observe that in 2\% of the cases, developers either preferred to rewrite the test ot completely disable it.

	\smallskip
	\item[Async Wait.] This flakiness is characterized by a test performing asynchronous calls without waiting properly for their result to become available. 
	As opposed to the \texttt{Concurrency} category presented above, which mostly relates to \emph{local} synchronization issues, this category usually refers to \emph{remote} resources being unavailable.
	For example, a test performing a request to a remote server and continuing the execution without ensuring the requested resource is available results in the test failing or passing depending on how quickly the resource becomes available. We confirm the results by Luo \etal~\cite{Luo:fse2014} on the high diffusion of this category, as it affects 52 of the validated tests (22\%). Moreover, the median effort is 3 (medium). The origin of the flakiness for all these tests is in test code. As for the fixing strategy, in most of the cases (86\%) developers fix it by means of a \texttt{waitFor} statement addition. 

	\smallskip
	\item[\underline{*Too Restrictive Range}.] This category was not included in the catalog by Luo \etal~\cite{Luo:fse2014}. In this category, some of the valid output values are outside the assertion range considered at test design time, so the test fails when they show up. In other words, such test cases have a range of predefined values for which the test is allowed to pass; if this range is defined too restrictively, tests may start failing in a not deterministic way. This type of flakiness additionally includes failures originating from improperly placed assertion statements causing the test to pass or fail independently of the test execution. In our dataset, 17\% of the flaky tests belong to this category, indicating that this problem is not uncommon. On the other hand, developers find this type of problem easy to fix (median effort=1) as it is always located in test code and only requires the proper definition of a range of values: indeed, in 65\% of the cases, developers address this problem by fixing or adjusting the range of values the assertion refers to. Interestingly, however, 16\% of the times the corresponding tests were disabled: likely, this is due to the limited time developers have to spend in understanding the correct range to use within the assertion and, thus, they prefer to keep having a green test suite. 

	\smallskip
	\item[Test Order Dependency.] The class is characterized by the result of the test run depending on the execution order of the tests. This occurs mostly due to tests not properly cleaning up after themselves (\eg restoring common states and shared variables) or failing to setup the necessary preconditions (\eg a test environment state) before starting to run~\cite{Luo:fse2014}. 11\% of the flaky tests are due to this reason, confirming its diffusion \cite{Palomba2017}. This is, by nature, a problem raising in test code. As for the effort, the median is equal to 2 (low)---thus, quite easy to fix for developers through the (i) modification of the output directory - so that the test uses a separate one, avoiding conflicts or (ii) checking the instance variables before the test is executed.

	\smallskip
	\item[\underline{*Test Case Timeout}.] Flaky tests experiencing non-deterministic timeouts related to a single test belong to this category. It is comparable to the Test Suite Timeout (reported later), with the difference that the size of a single test grew over time without adjusting the max runtime value. Various reasons (\eg failing to download prerequisites or a test not producing output for a fixed amount of time, which then is killed by the execution system assuming that the test stalled) can lead to the non-deterministic outcome. In our dataset, this flakiness type appears in 18 tests (8\%). Moreover, it is associated with a high effort to fix: despite it might seem that this problem would be only concerned with the increasing of the timeout, practitioners explained that finding the right timeout is unexpectedly hard. The increase of the timeout time is, obviously, the most frequent solution to this type of flakiness (85\%), however other fixing strategies such as (i) the addition of code that make the test faster or (ii) the extraction of a new test to reduce the risk of timeout are also sometimes taken into account. 

	\smallskip
	\item[Resource Leak.] Improper management of a external resources (\eg failing to release previously allocated memory or dereference a pointer) characterize this category. Additionally, this includes test failing due to garbage collection processes removing parts of the test execution environment or required resource (\eg a file stored in memory the test is reading multiple times)~\cite{Luo:fse2014}. This category represents the root-cause of 7\% of the flaky tests, and has a medium effort equals to 3. As for the origin, developers report that 15\% of them are due to production code issues: thus, we confirm that test flakiness is not just a problem of tests. With respect to the possible fixing strategies, in 50\% of the cases the test is modified so that the conflicting object is destroyed before continuing the execution. However, in the remaining cases, developers prefer to replace the test entirely or disable it: this result seems to confirm that dealing with test flakiness can be annoying for developers and that, in several cases, they prefer to skip the problem rather than find a solution to deal with it.
	\smallskip
	\item[\underline{*Platform Dependency.}] The Platform Dependency flakiness is: Non-deterministic test failures occurring only on specific platforms (\eg a test only failing in debug builds or on \textsc{32-bit Windows 7} systems). While a test failing consistently on a specific platform could represent a \emph{permanent} failure, developers consider such a test as flaky. In fact, if an organization employs cloud servers for builds/tests, machines with different characteristics (\eg Linux kernel version) could be provisioned according to available resources (in a seemingly random fashion from the user perspective), so the test fails intermittently.
	This type of flakiness occurs due to various reasons, such as a platform being unusually slow or missing preconditions (\eg the test failed to setup or download a required resource). They appear 10 times in our dataset and originate mostly in the test code (90\%), since they would most likely become permanent if they originated from a bug in the production code (as the tests would not run independently of what platform they are run on). The effort associated is 4 (high), as it turns to an infrastructure problem to be diagnosed and fixed. The high effort is also justified by the fixing strategies applied: indeed, 60\% of the times brand new platform-dependent tests are added, while in 40\% of the cases developers directly correct the existing code to avoid this flakiness.

	\smallskip
	\item[Float Precision.] As previously observed by Luo \etal~\cite{Luo:fse2014}, floating point operations may lead to non-deterministic test failures if potential precision over- and underflows are not considered (\eg converting a float formatted as a string, rounding or cutting to a certain number of significant digits). Additionally, this includes tests failures occurring due to different number of significant digits being reported in different executions. Even if the diffusion of this flakiness type is pretty low (only 6 cases), the reported effort is high, which indicates that developers might spend considerable time in dealing with this type of flaky tests. The fixing operation associated with this flakiness relates to the modification of the test to preserve precision.

	\smallskip
	\item[\underline{*Test Suite Timeout.}] Flaky tests originating because of the test suite non-deterministically timing out are classified as Test Suite Timeout flakiness. Test suites grow over time and the max runtime value is not always adjusted accordingly, leading to the test suites passing or failing depending on different random variables (\eg the network congestion, the execution speed of the platform or the type of build). Note that in a \texttt{Test Suite Timeout}, no single test is the cause of flakiness, rather the whole execution; this makes this cause different from \texttt{Test Case Timeout}, whose cause is a specific test.
	This flakiness cause appears very rarely (only 4 cases in our dataset) and the developers assigned a median effort score of 3.5. As this is test issue-related problem, its origin is within test files. To fix the flakiness, developers frequently decide to take actions able to fasten the execution of the test suite (75\% of the cases), while the remaining 25\% of times they perform an Extract Class refactoring \cite{Fowler:1999} to make the resulting suites more independent and faster.

	\smallskip
	\item[Time.] This cause of flakiness is characterized by tests relying on the executing system local time. For example, a test may fail due to the local system changing its day, \ie reporting a different day in the next iteration, or failing to take the timezone into account when comparing two timestamps. In line with Luo \etal~\cite{Luo:fse2014}, this kind of flakiness is generally not diffused; at the same time, we observe that it is always related to test code and is very easy to fix: the perceived easiness is due to the fact that in 75\% of the cases developers just disable the test, and thus only in a few cases (25\%) they actually try to preserve the time precision.

	\smallskip
	\item[Randomness.] The generation of random numbers may lead to non-deterministic failures if not all possible values generated including edge cases are considered (\eg the generation of zero when any number greater than zero was expected or the random number is used in a mathematical operation). We confirm the results of previous work~\cite{Luo:fse2014} on the very low diffusion of this flakiness cause. The associated effort is very low, the problem only appears in test files, and is always fixed by replacing the used random library call with a reliable number generator.

\end{description}

\noindent In our dataset, we do not find evidence of the presence of three out of the ten categories found by Luo \etal~\cite{Luo:fse2014}, namely \texttt{Network}, \texttt{I/O}, and \texttt{Unordered Collections}. On the one hand, this may be due to the type of analysis we do in this study: instead of classifying ourselves the likely causes behind test flakiness, we prefer asking to the original developers the type of flakiness occurring in the tests they fixed: in this way, we rely on the expertise of people that actually dealt with the investigated problems. On the other hand, the missing observation of those three categories might be due to the nature of the system we exploit, \ie the flakiness might manifest itself in different manners depending on the system taken into account. Our findings reveal the need for further research on the naturalness of flaky tests, where the role played by additional factors like project and organizational domains are assessed. At the same time, our taxonomy does not exclude the original one \cite{Luo:fse2014}, rather complements it. 

\roundedbox{
	We confirm the existence and frequencies of seven flakiness types revealed by Luo \etal~\cite{Luo:fse2014}. We discover four additional categories, three of which developers consider as the most effort-prone types of flakiness to deal with. Finally, we present the fixing strategies developers put in place to deal with flakiness and provided evidence that flaky tests can be also caused by problems in production code.
}

\section{\textbf{RQ$_2$} -- Flaky Tests: Relevance}
\label{sec:rq2}

\begin{figure}[htbp]
	\includegraphics[width=\columnwidth,keepaspectratio]{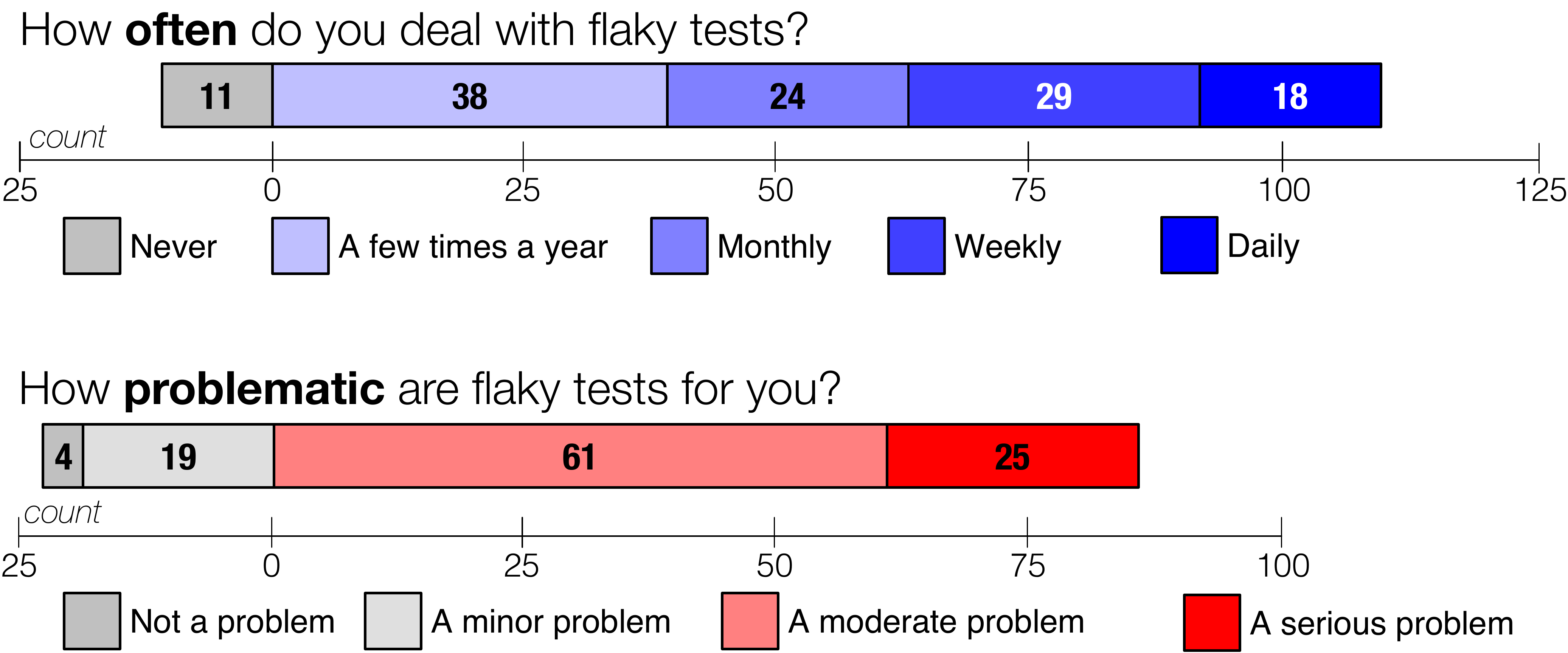}
	\caption{Frequency and relevance of the problem according to the respondents to our online survey.}
	\label{fig:relevance}
\end{figure}

Our second research question aims at understanding flaky tests prominence and how problematic they are according to developers. Figure~\ref{fig:relevance} shows the survey respondents' answers in terms of how frequently they deal with flaky tests and how problematic these tests are for those who deal with them at least a few times a year.

The vast majority (109 respondents) encounter the problem at least some times a year, and 58\% at least every month. Among these 109 respondents, 79\% find flaky tests to be at a moderate to serious problem, while 21\% a minor to not a problem, despite having to deal with them at least a few times per year.

The analysis of the open text comments received by the participants about the top problems they have with flaky tests reveal the presence of issues that can be grouped in three major themes:

\begin{description}[leftmargin=0.3cm]
	
	\item[Scheduling:] According to the opinions of 92 developers (77\%), flaky tests are \emph{time consuming} since reproducing the test failure is not easy and not always guaranteed to be possible. The debugging process includes multiple reruns of the same test while varying the level of log output and context taken into account. Additionally, since flaky tests fail intermittently, their priority is often lower than those of permanent failures, \ie developers need to find the time and request manager's permission to care for the flaky tests and the resources for the refactoring effort.
	
	\smallskip
	\item[Test Suite Reliability:] Once a test becomes flaky, 88 developers (73\%) report that \emph{the test is no longer fully reliable}. Thus, developers start to trust the test output less and, therefore, may start disregarding it, potentially leading to ignoring an actual failure. Moreover, the unexpectedness of flaky tests is an additional issue since it is unclear which tests will start to fail intermittently next.
	
	\smallskip
	\item[Developer Allocation:] 85 developers (71\%) mention that it is important to keep in mind that some flaky tests are more likely to occur (\eg tests relying on external dependencies) than others and, therefore, to distinguish how much attention shall be directed toward a specific type of failure. Since flaky tests are often intertwined with other tests, they require a certain \emph{level of knowledge to be able to fix them} and, thus, intermittent tests may lead to problems in the allocation of the available resources. 
	
\end{description}

\roundedbox{
	The collected developers' opinions indicate that flaky tests are rather frequent and a non-negligible problem, with possibly important consequences on resource allocation and scheduling, as well as on the reliability of the test suite.
}

\section{\textbf{RQ$_3$} -- Flaky Tests: Challenges}
\label{sec:rq3}

To answer what are the main challenges developers face when dealing with test flakiness, we further analyze the results of our online survey. When quoting developers, we refer to them by their ID, which corresponds to the incremental number assigned by the survey creation system.

\begin{table}[h]
	\caption{Pieces of information for fixing a flakiness (as emerged from the multivocal literature review), ranked by their importance (0--3) and difficulty (0--3) in obtaining, as rated by the survey respondents.}\label{tab:information}
	\begin{tabular}{|p{4.5cm}cc|}
		\hline
	\rowcolor{gray!50} \begin{tabular}[c]{@{}l@{}}Information\\ on the flaky test\end{tabular} & \begin{tabular}[c]{@{}l@{}}Importance\\ for fixing\end{tabular} & \begin{tabular}[c]{@{}l@{}}Difficulty\\ in obtaining\end{tabular} \\
	
	\textbf{\textsc{C$_f$}} - The context leading to failure & 2.69 & 1.65 \\
	\rowcolor{gray!20} \textbf{\textsc{N}} - The nature of the flakiness & 2.40 & 1.71 \\
	\textbf{\textsc{O}} - The origin of the flakiness & 2.22 & 1.17 \\
	\rowcolor{gray!20} \textbf{\textsc{E}} - The involved code elements & 2.21 & 1.13 \\
	\textbf{\textsc{C}} - The changes to perform the fix & 2.08 & 1.59 \\
	\rowcolor{gray!20} \textbf{\textsc{C$_p$}} - The context leading to passing & 1.95 & 1.20 \\
	\textbf{\textsc{Co}} - The commit introducing the flakiness & 1.89 & 0.75  \\
	\rowcolor{gray!20} \textbf{\textsc{H}} - The history of this test's flakiness (previous causes and fixes) & 1.79 & 0.83 \\\hline
	\end{tabular}
\end{table}

Table~\ref{tab:information} lists the eight pieces of information that emerged as potentially relevant for fixing a flaky behavior of a test from the multivocal literature review (Section~\ref{sec:survey_method}). As explained in the methdology, we asked participants to rate these pieces of information in terms of importance for fixing and difficulty in obtaining. Therefore, the rows are sorted according to the weighted average importance\footnote{We assigned a value of 0 (minimum) to answers stating `Not important at all' and of 3 (maximum) to answers stating `Extremely important', and 1 or 2 to the values in between.} according to our survey's respondents; the table also details the reported weighted average difficulty.\footnote{We assigned 0 (minimum) to both the values `Very easy' and `Easy', and 3 (maximum) to `Very difficult', and 1 or 2 to the values in between.} In this way, we identify pieces of information that developers actually perceive as challenging to obtain, thus indicating potential pain points to address with future research or better practices.

\smallskip
\noindent\textbf{\textbf{\textsc{C$_f$}}) Failing Context.} The first important and challenging piece of information to obtain is understanding what is the context that leads to the failing behavior. As P$_{161}$ puts it: ``the most difficult operation is reproducing a flaky test, as sometimes only 1/20 fails''. Additionally, P$_{174}$ reports: ``our UI tests are fairly slow, and the only real way to verify if the flakiness was resolved is to re-run the tests several times in the CI environment''. In other words, the verification of the failing behavior is even more complicated by some co-factors, such as the slowness of tests and the environment in which they are run. 

\smallskip
\noindent\textbf{\textbf{\textsc{N}}) Nature.} The second major challenge to the fixing process is the timely identification of the nature of the flakiness affecting a test. Understanding which of the causes discussed in \textbf{RQ$_2$} and in previous work \cite{Luo:fse2014} is the one leading to a flaky test represents a critical challenge for developers. Indeed, 52\% of the survey participants consider this information as `extremely important' and 37\% of them as `moderately important'. Similarly, 68\% of the developers report that figuring out the flakiness root-cause is `difficult' to `very difficult'. For instance, P$_{166}$ explains: ``a big challenge is to detect the root cause that leads to a flaky test. You need to check concurrency issues or cache related problems that might be common causes of flakiness''. This finding supports the need for empirical analyses aimed at further studying the phenomenon of test flakiness and the way it manifests itself. 

\smallskip
\noindent\textbf{\textbf{\textsc{E}}) Involved Code Elements.} Detecting the code elements involved in the flakiness is valuable information to obtain for 83\% of the developers, but most of them (75\%) consider it as not hard to gather. Therefore, this does not seem to be a major pain point, as the information is at hand by looking at the test code and log files. Nevertheless, there are still developers who consider this as major difficulty in the whole fixing process; looking more in depth into the reported opinions, we discover that the main factor characterizing the easiness of such information retrieval is how complex the source code actually is. As P$_{166}$ puts it: ``I think that for example finding the involved code might be rather easy if your code is well designed and written or might proof to be a complete nightmare if your codebase is a mess''. This may indicate that keeping test code quality high may naturally ease the flaky test fixing process.

\smallskip
\noindent\textbf{\textbf{\textsc{O}}) Origin.} The final potential challenge based on the information needs we identified in literature is whether the flakiness is caused by the test or by the production code. As shown in \textbf{RQ$_2$}, the flakiness can be even caused by problems in the production code and, apparently, having this information is important to speed-up the diagnosing of flaky tests. Indeed, 79\% of the participants evaluated this challenge as `moderately important' to `extremely important'. However, in this case, most of the developers (66\%) consider it pretty easy to obtain.

\smallskip
\noindent\textbf{\textbf{\textsc{C$_p$}},\textbf{\textsc{Co}},\textbf{\textsc{H}}) Passing Context, Commit, and History.} Finally, the context that leads to the passing behavior is considered only slightly or moderately important in 69\% of the cases because developers mainly aim at identifying the cases where behavior of the test leads to a failure. Similarly, the past flakiness history of a test seems to be poorly relevant during the fixing process for 72\% of the developers: This is likely due to the tests being subject to different flakiness types, meaning that the information from past history cannot be effectively reused to solve the current cause.

\begin{table}[htbp]
	\caption{Futher challenges due to flaky tests, as reported by the survey respondents.}
	\label{tab:results}
	\resizebox{\columnwidth}{!}{%
		\begin{tabular}{|llr|}
			\hline
			\rowcolor{gray!50} Category & Challenge & \# of mentions \\ \hline\hline
			\multirow{5}{*}{Design} 
			& Mocking Dependencies & 6 \\
			& \cellcolor{gray!20}Keeping Tests Decoupled & \cellcolor{gray!20}6 \\ 
			& Reliance on External Dependencies & 3 \\
			& \cellcolor{gray!20}Too Much Setup Code & \cellcolor{gray!20}3 \\   
			& Common Coding Style / Etiquette & 1 \\\hline\hline
			\multirow{4}{*}{Fixing} 
			& Uncertainty of Changes Fixing The Test & 7 \\
			& \cellcolor{gray!20}Not Detailed Enough Log & \cellcolor{gray!20}4 \\  
			& Lack of Insight Into The System & 4 \\  
			& \cellcolor{gray!20}Restructuring The Tests & \cellcolor{gray!20}2 \\\hline
		\end{tabular}
	}
\end{table}

Table \ref{tab:results} presents the additional challenges mentioned by the developers in the survey, by their reporting frequency. 
Overall, the first observation is related to the development phases mentioned by the survey participants. While we expect to find critical challenges arising during the diagnosing and fixing process, the fact that developers report the presence of challenges at design-time is somehow surprising and unexpected. OLur findings indicate how the problem of flakiness is \emph{spread over the entire life-cycle of tests}.

\smallskip
\textbf{Design.} According to our participants, the probability for a test to be flaky can be reduced if good design principles are applied while developing it, thus confirming previous findings in the field \cite{palomba2017does}. Six developers report that keeping tests decoupled and mocking dependencies are major challenges when designing test to be less prone to flakiness. In the former case, keeping low the coupling between tests is a key property because there is the risk that, quoting P$_{130}$, ``one test sets something up that affects another test, and so on''. In the second case, understanding when it is convenient to mock external objects can be useful to ``control the environment (making sure packages installed are all the right version, mocking time, mocking/stubbing out dependencies on external services)'' [P$_{80}$]. 
Additional challenges are represented by \begin{inparaenum}[(i)]
	\item designing systems that do not rely too much on external dependencies, to reduce the risk of intermittent failures due to not controllable sources;
	\item reducing the amount of setup code required, so that flakiness eventually due to problems with the setup of tests can be more easily diagnosed;
	and
	\item establishing a common coding style and etiquette, to ease the understanding of test code.\end{inparaenum}

\smallskip
\textbf{Fixing.} Seven participants report as a main challenge the uncertainty of not knowing whether a code change actually fixes the flakiness (it may disappear by chance). This is currently solved by manually re-running multiple times the supposedly fixed flaky tests, but it is not considered as an optimal solution. Finally, log files with insufficient levels of detail and a lack of insights into the system are also reported as a major challenge, as well as the process of restructuring the test to remedy the flakiness.

\roundedbox{
	Reproducing the context leading to the test failure and understanding the nature of the flakiness are the most important, yet challenging needs according to developers. Moreover, designing test code properly to avoid flakiness emerged as an additional challenge not mentioned in the reviewed academic and gray literature.
}

\section{Discussion and Implications}
\label{sec:discussion}

Our results highlighted a number of points to be further discussed and several implications for the research community:

\smallskip
\textbf{Flaky tests as a relevant research problem.} Besides confirming seven of the root-causes behind flaky tests originally defined by Luo \etal \cite{Luo:fse2014}, our work let emerge the existence of four additional types of flakiness previously unknown, \ie \texttt{Too Restrictive Range}, \texttt{Test Case Timeout}, \texttt{Platform Dependency}, and \texttt{Test Suite Timeout}. We found them to be highly diffused over our dataset, and the latter three causes are those that developers associated with a higher fixing effort. This finding calls for further research in this area to both understand \emph{how} such flaky tests are introduced and managed, as well as devise automatic solutions to locate and fix them. The worrisome finding that, in a not negligible amount of cases, developers prefer to just disable flaky tests rather than properly deal with them is a urgent call for researchers to find solutions to help practitioners dealing with developers' information needs (as emerged from this study) to reduce their efforts.

\smallskip
\textbf{An organizational view.} The presence of flaky tests leads to several undesired problems and consequences connected to organizational aspects (\eg resource allocation issues). This finding highlights the current inability of development teams to properly manage flaky tests; thus, we argue that a broader view of test flakiness should not only involve source code and its management, but also how flaky tests impact teams' organization and how their resolution process may be optimized. For instance, research can be conducted to investigate approaches able to recommend \emph{who} should fix a flaky test based on developers' knowledge/experience, flakiness type, and the available time.

\smallskip
\textbf{Preventing Flaky Tests.} As test flakiness is problematic, it is of the greatest importance the definition of methods able to \emph{prevent} the introduction of flaky tests. We envision the adoption of machine learning techniques---properly trained on the features characterizing each flakiness type---to be a promising way to help developers in preventively spotting tests that have the highest risk to become flaky in the future. 

\smallskip
\textbf{Design for Testability.} Our participants revealed that finding how to design a test code to avoid flakiness is an important challenge to face. This motivates the growing research area around test code quality~\cite{fraser2013whole,palomba2016automatic,palomba2018automatic,spadini2017mock,spadini2018relation,spadini2019mock,spadini2019test} and provides two promising directions that the research community can focus on: (i) the definition of a set of \emph{design patterns} that can support the creation of deterministic tests; (ii) the definition of a set of flakiness-related \emph{anti-patterns} that practitioners should avoid when writing test cases. While some initial steps have been done about the relation between test smells and flaky tests~\cite{palomba2017does,palombaEMSE2019}, further investigation is necessary.

\smallskip
\textbf{Flaky Test Prioritization.} The results on the effort required to fix each flaky test type, as well as their perceived importance, indicated that not all flaky tests are equally problematic. As large codebases might contain a higher number of tests suffering flakiness issues, approaches that can rank these tests by exploiting effort-related information would help developers in maintaining high quality test suites. Machine learning is a promising approach to exploit to accomplish this task, as it could spot patterns in the code that are the fingerprints of certain types of flakiness.

\section{Threats to Validity}
\label{sec:threats}

This section discusses the threats that might have influenced our findings, despite our mitigation strategies. 

\smallskip
\textbf{Construct Validity.} Threats in this category are mainly concerned to the way we built the dataset of flaky tests (\textbf{RQ$_1$}), the relevance of the problem (\textbf{RQ$_2$}), and the potential challenges (\textbf{RQ$_3$}).
To obtain a reliable analysis of the flaky test types, we directly involved the professional \textsc{Mozilla} developers and asked them to analyze the flakiness cases they fixed themselves. Although we considered flaky tests they fixed in a recent time and provided them with the links to their own patches as well as issue reports, we cannot exclude that they could not recall the required fixing effort precisely.
To collect information about the frequency and relevance of flakiness for developers, we used an online survey. Although we spread the survey through several channels, our survey responses may suffer from a self-selection or voluntary response bias: People who volunteered to respond may be more involved with flakiness than the average developer. To mitigate this bias, we introduced a donation-based incentive of 2 USD to a charity per valid respondent. Although our results concerning the frequency are aligned with those reported by Luo \etal~\cite{Luo:fse2014}, it is still possible that we have more extreme results about flakiness frequency and harmfulness than what would be in the average practice.
Finally, besides inquiring developers on the major challenges they face when dealing with test flakiness, in \textbf{RQ$_3$} we came up with a set of information needs anecdotally considered relevant by practitioners. We did this by surveying both white and gray literature following the validated guidelines by Garousi \etal \cite{garousi2016need}. However, we cannot exclude that other sources presented further potential information needs and challenges when dealing with test flakiness.

\smallskip
\textbf{Conclusion Validity.} As for the analysis methods exploited in our study, in \textbf{RQ$_1$} we computed statistics on the answers provided by the participants of our survey. In the context of \textbf{RQ$_2$} we defined a taxonomy of flaky test types by relying on an iterative content analysis that involved three software engineering researchers having experience with flaky tests. Although such iterative process is designed to mitigate the possibility of classification errors, subjective judgement may have played a role in the elaboration of the themes. A similar process has been done in the context of \textbf{RQ$_3$} to classify the challenges emerged from the multivocal literature review. We cannot exclude human errors or that some important challenges, not present in the sources examined, were not reported in the survey. Further replications and analyses would be beneficial to corroborate our findings.

\smallskip
\textbf{External validity.} Threats related to the generalizability of our findings concern size and nature of the dataset exploited. We considered 200 flaky tests coming from \textsc{Mozilla}. Although the company has been the subject of several software engineering studies, which later also generalized to other contexts, the taxonomy we created may (and may not) be applicable to other systems and, therefore, further studies are needed to investigate this aspect.

\section{Related Work}
\label{sec:related}

A number of researchers and practitioners reported that some flaky tests can be a serious problem in automated regression testing~\cite{palomba2017does,Luo:fse2014, Bell:2014, Fowler:flakyTests, Memon:2013, Muslu:2011, Zhang:2014:ERT}.
Memon and Cohen~\cite{Memon:2013} described multiple negative effects that flaky tests can create, discovering that their occurrence may even lead to features not being included in a release as they were not able to be tested sufficiently beforehand~\cite{Memon:2013}. 
Marinescu \etal \cite{Marinescu:2014:CFA} and Hilton \etal \cite{coverageASE} analyzed the evolution of test suite coverage, reporting that the presence of flaky tests produces an intermittent variation of the branch coverage.

Other researchers tried to understand the reasons behind test code flakiness. 
Luo \etal \cite{Luo:fse2014} manually analyzed the source code of tests involved in 201 commits that likely fixed flaky tests, defining a taxonomy of ten common root-causes. Moreover, they also provide hints on how developers usually fix flaky tests.
As a result, they found that the top three common causes of flakiness are related to asynchronous wait, concurrency, and test order dependency issues.  
While the taxonomy of flaky test root causes built in the context of \textbf{RQ$_1$} is mostly aligned with the one presented by Luo \etal \cite{Luo:fse2014}, our study reveals the existence of additional causes for flakiness. In this sense, our paper can be seen as an additional source aimed at better understanding the phenomenon of flaky tests.

Also other researchers investigated the motivations behind flaky tests as well as devised strategies for their automatic identification. 
For instance, Palomba and Zaidman \cite{palomba2017does,palombaEMSE2019} discovered that test smells \cite{Deursen01refactoringtest} may induce test code flakiness; moreover, they found that the removal of such smells through refactoring also induce fixing flaky tests.
Zhang \etal \cite{Zhang:2014:ERT} focused instead on test suites affected by test dependency issues, reporting methodologies to identify these tests. Muslu \etal \cite{Muslu:2011} found that test isolation may help in fault localization, while Bell and Kaiser \cite{Bell:2014} proposed a technique able to isolate test cases in Java applications by tracking the shared objects in memory. Bell \etal \cite{bell2018deflaker} proposed \textsc{DeFlaker}, an automated technique that identifies flaky tests by running a mix of static and dynamic analyses.

Another well-studied root cause of flaky test is concurrency. 
Farchi \etal \cite{Farchi:2003} identified a set of common erroneous patterns in concurrent code, while Lu \etal \cite{Lu:2008:LMC} reported a comprehensive study into the characteristics of concurrency bugs. Jin \etal \cite{Jin:2011} devised a technique for automatically fixing concurrency bugs by analyzing the single-variable atomicity violations.
With respect to the studies discussed above, our work can be considered as complementary as it provides, for the first time, an in-depth investigation of the developer's perspective, which includes the nature of flaky tests and fixing efforts, the relevance of the problem, and the most important challenges that face when handling flaky tests.

\section{Conclusion}
\label{sec:conclusion}

We presented an empirical study aimed at improving our scientific knowledge around the problem of test flakiness. We first assembled a dataset of 200 flaky tests, asking the original developers that fixed the problem to analyze them in terms of nature, fixing effort, and origin of the flakiness. Then, we used this annotated dataset (which we also make publicly available) to define a taxonomy of flaky test types, which comprises seven categories already known in literature \cite{Luo:fse2014} as well as four additional ones; out of these four, we found that three are the most demanding overall in terms of fixing efforts. In parallel, we performed a survey study having the goal of collecting developers' opinions on the relevance of flakiness in practice, as well as major challenges and information needs they have when dealing with flaky tests. This study revealed a set of problems and needs that the research community should carefully look at as to improve the way test flakiness is managed by practitioners.

\begin{enumerate}

\item Evidence on four new categories of flaky tests---important for both researchers (called to devise techniques to deal with these new types) and practitioners (called to fix these issues);

\smallskip
\item First reported evidence that flakiness is also caused by problems in production code;

\smallskip
\item Surprising evidence from our analyses on the effort required to fix flaky tests, like the timeout problem;

\smallskip
\item Evidence on unexpected side-effects of flaky tests, such as the issues they induce to organizational aspects (e.g., resources allocation). This gives insights to practitioners to reflect on their practices with flaky tests.

\end{enumerate}	

It is our hope that the insights we have discovered lead to more improved techniques and tools, based on research, to aid developers handle flaky tests.

\begin{acks}
	The authors would like to thank all the \mozilla developers who participated in the creation of the dataset, as well as the respondents to our survey: You provided us and our research community with very precious data.
	Bacchelli and Palomba gratefully acknowledge the support of the Swiss National Science Foundation through the SNF Project No. PP00P2\_170529.
\end{acks}
\balance

\newpage

\bibliographystyle{ACM-Reference-Format}
\bibliography{main}


\begin{thebibliography}{42}


\ifx \showCODEN    \undefined \def \showCODEN     #1{\unskip}     \fi
\ifx \showDOI      \undefined \def \showDOI       #1{#1}\fi
\ifx \showISBNx    \undefined \def \showISBNx     #1{\unskip}     \fi
\ifx \showISBNxiii \undefined \def \showISBNxiii  #1{\unskip}     \fi
\ifx \showISSN     \undefined \def \showISSN      #1{\unskip}     \fi
\ifx \showLCCN     \undefined \def \showLCCN      #1{\unskip}     \fi
\ifx \shownote     \undefined \def \shownote      #1{#1}          \fi
\ifx \showarticletitle \undefined \def \showarticletitle #1{#1}   \fi
\ifx \showURL      \undefined \def \showURL       {\relax}        \fi
\providecommand\bibfield[2]{#2}
\providecommand\bibinfo[2]{#2}
\providecommand\natexlab[1]{#1}
\providecommand\showeprint[2][]{arXiv:#2}

\bibitem[\protect\citeauthoryear{??}{sur}{2019}]%
        {surveygizmo}
 \bibinfo{year}{2019}\natexlab{}.
\newblock \bibinfo{title}{{SurveyGizmo}}.
\newblock \bibinfo{howpublished}{\url{https://www.surveygizmo.com}}.
\newblock


\bibitem[\protect\citeauthoryear{2018.}{2018.}{2018a}]%
        {fse}
\bibfield{author}{\bibinfo{person}{2018.}} \bibinfo{year}{2018}\natexlab{a}.
\newblock \bibinfo{title}{{ACM Joint European Software Engineering Conference
  and Symposium on the Foundations of Software Engineering.}}
\newblock
\newblock
\urldef\tempurl%
\url{http://www.esec-fse.org}
\showURL{%
\tempurl}


\bibitem[\protect\citeauthoryear{2018.}{2018.}{2018b}]%
        {issta}
\bibfield{author}{\bibinfo{person}{2018.}} \bibinfo{year}{2018}\natexlab{b}.
\newblock \bibinfo{title}{{ACM SIGSOFT International Symposium on Software
  Testing and Analysis.}}
\newblock
\newblock
\urldef\tempurl%
\url{https://conf.researchr.org/series/issta}
\showURL{%
\tempurl}


\bibitem[\protect\citeauthoryear{2018.}{2018.}{2018c}]%
        {tosem}
\bibfield{author}{\bibinfo{person}{2018.}} \bibinfo{year}{2018}\natexlab{c}.
\newblock \bibinfo{title}{{ACM Transactions on Software Engineering and
  Methodology.}}
\newblock
\newblock
\urldef\tempurl%
\url{https://tosem.acm.org}
\showURL{%
\tempurl}


\bibitem[\protect\citeauthoryear{2018.}{2018.}{2018d}]%
        {icsme}
\bibfield{author}{\bibinfo{person}{2018.}} \bibinfo{year}{2018}\natexlab{d}.
\newblock \bibinfo{title}{{IEEE TCSE International Conference on Software
  Maintenance and Evolution.}}
\newblock
\newblock
\urldef\tempurl%
\url{http://conferences.computer.org/icsm/}
\showURL{%
\tempurl}


\bibitem[\protect\citeauthoryear{2018.}{2018.}{2018e}]%
        {tse}
\bibfield{author}{\bibinfo{person}{2018.}} \bibinfo{year}{2018}\natexlab{e}.
\newblock \bibinfo{title}{{IEEE Transactions on Software Engineering.}}
\newblock
\newblock
\urldef\tempurl%
\url{https://www.computer.org/web/tse}
\showURL{%
\tempurl}


\bibitem[\protect\citeauthoryear{2018.}{2018.}{2018f}]%
        {icse}
\bibfield{author}{\bibinfo{person}{2018.}} \bibinfo{year}{2018}\natexlab{f}.
\newblock \bibinfo{title}{{IEEE/ACM International Conference on Software
  Engineering.}}
\newblock
\newblock
\urldef\tempurl%
\url{http://www.icse-conferences.org}
\showURL{%
\tempurl}


\bibitem[\protect\citeauthoryear{2018.}{2018.}{2018g}]%
        {icst}
\bibfield{author}{\bibinfo{person}{2018.}} \bibinfo{year}{2018}\natexlab{g}.
\newblock \bibinfo{title}{{International Conference on Software Testing.}}
\newblock
\newblock
\urldef\tempurl%
\url{https://www.es.mdh.se/icst2018/}
\showURL{%
\tempurl}


\bibitem[\protect\citeauthoryear{2018.}{2018.}{2018h}]%
        {emse}
\bibfield{author}{\bibinfo{person}{2018.}} \bibinfo{year}{2018}\natexlab{h}.
\newblock \bibinfo{title}{{Springer's Empirical Software Engineering Journal.}}
\newblock
\newblock
\urldef\tempurl%
\url{https://link.springer.com/journal/10664}
\showURL{%
\tempurl}


\bibitem[\protect\citeauthoryear{Bell and Kaiser}{Bell and Kaiser}{2014}]%
        {Bell:2014}
\bibfield{author}{\bibinfo{person}{Jonathan Bell} {and} \bibinfo{person}{Gail
  Kaiser}.} \bibinfo{year}{2014}\natexlab{}.
\newblock \showarticletitle{Unit Test Virtualization with {VMVM}}. In
  \bibinfo{booktitle}{\emph{Proceedings of the International Conference on
  Software Engineering (ICSE)}}. \bibinfo{publisher}{ACM},
  \bibinfo{pages}{550--561}.
\newblock
\showISBNx{978-1-4503-2756-5}
\urldef\tempurl%
\url{https://doi.org/10.1145/2568225.2568248}
\showDOI{\tempurl}


\bibitem[\protect\citeauthoryear{Bell, Legunsen, Hilton, Eloussi, Yung, and
  Marinov}{Bell et~al\mbox{.}}{2018}]%
        {bell2018deflaker}
\bibfield{author}{\bibinfo{person}{Jonathan Bell}, \bibinfo{person}{Owolabi
  Legunsen}, \bibinfo{person}{Michael Hilton}, \bibinfo{person}{Lamyaa
  Eloussi}, \bibinfo{person}{Tifany Yung}, {and} \bibinfo{person}{Darko
  Marinov}.} \bibinfo{year}{2018}\natexlab{}.
\newblock \showarticletitle{DeFlaker: Automatically Detecting Flaky Tests}. In
  \bibinfo{booktitle}{\emph{Proceedings of the International Conference on
  Software Engineering (ICSE)}}.
\newblock
\newblock
\shownote{To Appear.}


\bibitem[\protect\citeauthoryear{Corbin and Strauss}{Corbin and
  Strauss}{1990}]%
        {corbin1990grounded}
\bibfield{author}{\bibinfo{person}{Juliet~M Corbin} {and}
  \bibinfo{person}{Anselm Strauss}.} \bibinfo{year}{1990}\natexlab{}.
\newblock \showarticletitle{Grounded theory research: Procedures, canons, and
  evaluative criteria}.
\newblock \bibinfo{journal}{\emph{Qualitative sociology}} \bibinfo{volume}{13},
  \bibinfo{number}{1} (\bibinfo{year}{1990}), \bibinfo{pages}{3--21}.
\newblock


\bibitem[\protect\citeauthoryear{Eck, Palomba, Castelluccio, and Bacchelli}{Eck
  et~al\mbox{.}}{2019}]%
        {eck:appendix}
\bibfield{author}{\bibinfo{person}{Moritz Eck}, \bibinfo{person}{Fabio
  Palomba}, \bibinfo{person}{Marco Castelluccio}, {and}
  \bibinfo{person}{Alberto Bacchelli}.} \bibinfo{year}{2019}\natexlab{}.
\newblock \bibinfo{title}{Data and materials for: `Understanding Flaky Tests:
  The Developer's Perspective'}.
\newblock \bibinfo{howpublished}{\url{https://doi.org/10.5281/zenodo.3265830}}.
\newblock


\bibitem[\protect\citeauthoryear{Farchi, Nir, and Ur}{Farchi
  et~al\mbox{.}}{2003}]%
        {Farchi:2003}
\bibfield{author}{\bibinfo{person}{E. Farchi}, \bibinfo{person}{Y. Nir}, {and}
  \bibinfo{person}{S. Ur}.} \bibinfo{year}{2003}\natexlab{}.
\newblock \showarticletitle{Concurrent bug patterns and how to test them}. In
  \bibinfo{booktitle}{\emph{Proceedings International Parallel and Distributed
  Processing Symposium}}. \bibinfo{pages}{7 pp.--}.
\newblock
\showISSN{1530-2075}
\urldef\tempurl%
\url{https://doi.org/10.1109/IPDPS.2003.1213511}
\showDOI{\tempurl}


\bibitem[\protect\citeauthoryear{Flanigan, McFarlane, and Cook}{Flanigan
  et~al\mbox{.}}{2008}]%
        {flanigan2008conducting}
\bibfield{author}{\bibinfo{person}{Timothy~S Flanigan}, \bibinfo{person}{Emily
  McFarlane}, {and} \bibinfo{person}{Sarah Cook}.}
  \bibinfo{year}{2008}\natexlab{}.
\newblock \showarticletitle{Conducting survey research among physicians and
  other medical professionals: a review of current literature}. In
  \bibinfo{booktitle}{\emph{Proceedings of the Survey Research Methods Section,
  American Statistical Association}}, Vol.~\bibinfo{volume}{1}.
  \bibinfo{pages}{4136--47}.
\newblock


\bibitem[\protect\citeauthoryear{Fowler}{Fowler}{[n. d.]}]%
        {Fowler:flakyTests}
\bibfield{author}{\bibinfo{person}{Martin Fowler}.} \bibinfo{year}{[n.
  d.]}\natexlab{}.
\newblock \bibinfo{title}{Eradicating non-determinism in tests.}
\newblock
\newblock
\urldef\tempurl%
\url{https://martinfowler.com/articles/nonDeterminism.html}
\showURL{%
\tempurl}


\bibitem[\protect\citeauthoryear{Fowler}{Fowler}{1999}]%
        {Fowler:1999}
\bibfield{author}{\bibinfo{person}{M. Fowler}.}
  \bibinfo{year}{1999}\natexlab{}.
\newblock \bibinfo{booktitle}{\emph{Refactoring: improving the design of
  existing code}}.
\newblock \bibinfo{publisher}{Addison-Wesley}.
\newblock


\bibitem[\protect\citeauthoryear{Fraser and Arcuri}{Fraser and Arcuri}{2013}]%
        {fraser2013whole}
\bibfield{author}{\bibinfo{person}{Gordon Fraser} {and} \bibinfo{person}{Andrea
  Arcuri}.} \bibinfo{year}{2013}\natexlab{}.
\newblock \showarticletitle{Whole test suite generation}.
\newblock \bibinfo{journal}{\emph{IEEE Transactions on Software Engineering}}
  \bibinfo{volume}{39}, \bibinfo{number}{2} (\bibinfo{year}{2013}),
  \bibinfo{pages}{276--291}.
\newblock


\bibitem[\protect\citeauthoryear{Garousi, Felderer, and
  M{\"a}ntyl{\"a}}{Garousi et~al\mbox{.}}{2016}]%
        {garousi2016need}
\bibfield{author}{\bibinfo{person}{Vahid Garousi}, \bibinfo{person}{Michael
  Felderer}, {and} \bibinfo{person}{Mika~V M{\"a}ntyl{\"a}}.}
  \bibinfo{year}{2016}\natexlab{}.
\newblock \showarticletitle{The need for multivocal literature reviews in
  software engineering: complementing systematic literature reviews with grey
  literature}. In \bibinfo{booktitle}{\emph{Proceedings of the 20th
  International Conference on Evaluation and Assessment in Software
  Engineering}}. ACM, \bibinfo{pages}{26}.
\newblock


\bibitem[\protect\citeauthoryear{Hilton, Bell, and Marinov}{Hilton
  et~al\mbox{.}}{2018}]%
        {coverageASE}
\bibfield{author}{\bibinfo{person}{Michael Hilton}, \bibinfo{person}{Jonathan
  Bell}, {and} \bibinfo{person}{Darko Marinov}.}
  \bibinfo{year}{2018}\natexlab{}.
\newblock \showarticletitle{A Large-Scale, Longitudinal Study of Test Coverage
  Evolution}. In \bibinfo{booktitle}{\emph{33rd IEEE/ACM International
  Conference on Automated Software Engineering}} \emph{(\bibinfo{series}{ASE
  2018})}.
\newblock
\urldef\tempurl%
\url{http://jonbell.net/publications/coverage}
\showURL{%
\tempurl}


\bibitem[\protect\citeauthoryear{Jin, Song, Zhang, Lu, and Liblit}{Jin
  et~al\mbox{.}}{2011}]%
        {Jin:2011}
\bibfield{author}{\bibinfo{person}{Guoliang Jin}, \bibinfo{person}{Linhai
  Song}, \bibinfo{person}{Wei Zhang}, \bibinfo{person}{Shan Lu}, {and}
  \bibinfo{person}{Ben Liblit}.} \bibinfo{year}{2011}\natexlab{}.
\newblock \showarticletitle{Automated Atomicity-violation Fixing}. In
  \bibinfo{booktitle}{\emph{Proceedings of the 32Nd ACM SIGPLAN Conference on
  Programming Language Design and Implementation (PLDI)}}.
  \bibinfo{publisher}{ACM}, \bibinfo{pages}{389--400}.
\newblock
\showISBNx{978-1-4503-0663-8}
\urldef\tempurl%
\url{https://doi.org/10.1145/1993498.1993544}
\showDOI{\tempurl}


\bibitem[\protect\citeauthoryear{Johnson and Onwuegbuzie}{Johnson and
  Onwuegbuzie}{2004}]%
        {johnson2004mixed}
\bibfield{author}{\bibinfo{person}{R~Burke Johnson} {and}
  \bibinfo{person}{Anthony~J Onwuegbuzie}.} \bibinfo{year}{2004}\natexlab{}.
\newblock \showarticletitle{Mixed methods research: A research paradigm whose
  time has come}.
\newblock \bibinfo{journal}{\emph{Educational researcher}}
  \bibinfo{volume}{33}, \bibinfo{number}{7} (\bibinfo{year}{2004}),
  \bibinfo{pages}{14--26}.
\newblock


\bibitem[\protect\citeauthoryear{Lu, Park, Seo, and Zhou}{Lu
  et~al\mbox{.}}{2008}]%
        {Lu:2008:LMC}
\bibfield{author}{\bibinfo{person}{Shan Lu}, \bibinfo{person}{Soyeon Park},
  \bibinfo{person}{Eunsoo Seo}, {and} \bibinfo{person}{Yuanyuan Zhou}.}
  \bibinfo{year}{2008}\natexlab{}.
\newblock \showarticletitle{Learning from Mistakes: A Comprehensive Study on
  Real World Concurrency Bug Characteristics}. In
  \bibinfo{booktitle}{\emph{Proceedings of the International Conference on
  Architectural Support for Programming Languages and Operating Systems
  (ASPLOS)}}. \bibinfo{publisher}{ACM}, \bibinfo{pages}{329--339}.
\newblock
\showISBNx{978-1-59593-958-6}
\urldef\tempurl%
\url{https://doi.org/10.1145/1346281.1346323}
\showDOI{\tempurl}


\bibitem[\protect\citeauthoryear{Luo, Hariri, Eloussi, and Marinov}{Luo
  et~al\mbox{.}}{2014}]%
        {Luo:fse2014}
\bibfield{author}{\bibinfo{person}{Qingzhou Luo}, \bibinfo{person}{Farah
  Hariri}, \bibinfo{person}{Lamyaa Eloussi}, {and} \bibinfo{person}{Darko
  Marinov}.} \bibinfo{year}{2014}\natexlab{}.
\newblock \showarticletitle{An Empirical Analysis of Flaky Tests}. In
  \bibinfo{booktitle}{\emph{Proceedings of the SIGSOFT International Symposium
  on Foundations of Software Engineering (FSE)}}. \bibinfo{publisher}{ACM},
  \bibinfo{pages}{643--653}.
\newblock
\showISBNx{978-1-4503-3056-5}
\urldef\tempurl%
\url{https://doi.org/10.1145/2635868.2635920}
\showDOI{\tempurl}


\bibitem[\protect\citeauthoryear{Marinescu, Hosek, and Cadar}{Marinescu
  et~al\mbox{.}}{2014}]%
        {Marinescu:2014:CFA}
\bibfield{author}{\bibinfo{person}{Paul Marinescu}, \bibinfo{person}{Petr
  Hosek}, {and} \bibinfo{person}{Cristian Cadar}.}
  \bibinfo{year}{2014}\natexlab{}.
\newblock \showarticletitle{Covrig: A Framework for the Analysis of Code, Test,
  and Coverage Evolution in Real Software}. In
  \bibinfo{booktitle}{\emph{Proceedings of the International Symposium on
  Software Testing and Analysis (ISSTA)}}. \bibinfo{publisher}{ACM},
  \bibinfo{pages}{93--104}.
\newblock
\showISBNx{978-1-4503-2645-2}
\urldef\tempurl%
\url{https://doi.org/10.1145/2610384.2610419}
\showDOI{\tempurl}


\bibitem[\protect\citeauthoryear{Memon and Cohen}{Memon and Cohen}{2013}]%
        {Memon:2013}
\bibfield{author}{\bibinfo{person}{Atif~M. Memon} {and}
  \bibinfo{person}{Myra~B. Cohen}.} \bibinfo{year}{2013}\natexlab{}.
\newblock \showarticletitle{Automated Testing of GUI Applications: Models,
  Tools, and Controlling Flakiness}. In \bibinfo{booktitle}{\emph{Proceedings
  of the International Conference on Software Engineering (ICSE)}}.
  \bibinfo{publisher}{IEEE}, \bibinfo{pages}{1479--1480}.
\newblock
\showISBNx{978-1-4673-3076-3}


\bibitem[\protect\citeauthoryear{Mu\c{s}lu, Soran, and Wuttke}{Mu\c{s}lu
  et~al\mbox{.}}{2011}]%
        {Muslu:2011}
\bibfield{author}{\bibinfo{person}{Kivan\c{c} Mu\c{s}lu},
  \bibinfo{person}{Bilge Soran}, {and} \bibinfo{person}{Jochen Wuttke}.}
  \bibinfo{year}{2011}\natexlab{}.
\newblock \showarticletitle{Finding Bugs by Isolating Unit Tests}. In
  \bibinfo{booktitle}{\emph{Proceedings of the SIGSOFT Symposium on Foundations
  of Software Engineering and the European Conference on Software Engineering
  (ESEC/FSE)}}. \bibinfo{publisher}{ACM}, \bibinfo{pages}{496--499}.
\newblock
\showISBNx{978-1-4503-0443-6}
\urldef\tempurl%
\url{https://doi.org/10.1145/2025113.2025202}
\showDOI{\tempurl}


\bibitem[\protect\citeauthoryear{Oppenheim}{Oppenheim}{1992}]%
        {Oppenheim:1992}
\bibfield{author}{\bibinfo{person}{A.~N. Oppenheim}.}
  \bibinfo{year}{1992}\natexlab{}.
\newblock \bibinfo{booktitle}{\emph{Questionnaire Design, Interviewing and
  Attitude Measurement}}.
\newblock \bibinfo{publisher}{Pinter Publishers}.
\newblock


\bibitem[\protect\citeauthoryear{Palomba, Panichella, Zaidman, Oliveto, and
  De~Lucia}{Palomba et~al\mbox{.}}{2016}]%
        {palomba2016automatic}
\bibfield{author}{\bibinfo{person}{Fabio Palomba}, \bibinfo{person}{Annibale
  Panichella}, \bibinfo{person}{Andy Zaidman}, \bibinfo{person}{Rocco Oliveto},
  {and} \bibinfo{person}{Andrea De~Lucia}.} \bibinfo{year}{2016}\natexlab{}.
\newblock \showarticletitle{Automatic test case generation: What if test code
  quality matters?}. In \bibinfo{booktitle}{\emph{Proceedings of the 25th
  International Symposium on Software Testing and Analysis}}. ACM,
  \bibinfo{pages}{130--141}.
\newblock


\bibitem[\protect\citeauthoryear{Palomba and Zaidman}{Palomba and
  Zaidman}{2017a}]%
        {Palomba2017}
\bibfield{author}{\bibinfo{person}{Fabio Palomba} {and} \bibinfo{person}{Andy
  Zaidman}.} \bibinfo{year}{2017}\natexlab{a}.
\newblock \showarticletitle{{Does refactoring of test smells induce fixing
  flaky tests?}}. In \bibinfo{booktitle}{\emph{Proceedings - 2017 IEEE
  International Conference on Software Maintenance and Evolution, ICSME 2017}}.
  \bibinfo{pages}{1--12}.
\newblock
\showISBNx{9781538609927}
\urldef\tempurl%
\url{https://doi.org/10.1109/ICSME.2017.12}
\showDOI{\tempurl}


\bibitem[\protect\citeauthoryear{Palomba and Zaidman}{Palomba and
  Zaidman}{2017b}]%
        {palomba2017does}
\bibfield{author}{\bibinfo{person}{Fabio Palomba} {and} \bibinfo{person}{Andy
  Zaidman}.} \bibinfo{year}{2017}\natexlab{b}.
\newblock \showarticletitle{Does refactoring of test smells induce fixing flaky
  tests?}. In \bibinfo{booktitle}{\emph{Software Maintenance and Evolution
  (ICSME), 2017 IEEE International Conference on}}. IEEE,
  \bibinfo{pages}{1--12}.
\newblock


\bibitem[\protect\citeauthoryear{Palomba and Zaidman}{Palomba and
  Zaidman}{2019}]%
        {palombaEMSE2019}
\bibfield{author}{\bibinfo{person}{Fabio Palomba} {and} \bibinfo{person}{Andy
  Zaidman}.} \bibinfo{year}{2019}\natexlab{}.
\newblock \showarticletitle{The smell of fear: On the relation between test
  smells and flaky tests}.
\newblock \bibinfo{journal}{\emph{Journal of Empirical Software Engineering}}
  (\bibinfo{year}{2019}).
\newblock


\bibitem[\protect\citeauthoryear{Palomba, Zaidman, and Lucia}{Palomba
  et~al\mbox{.}}{2018}]%
        {palomba2018automatic}
\bibfield{author}{\bibinfo{person}{Fabio Palomba}, \bibinfo{person}{Andy
  Zaidman}, {and} \bibinfo{person}{AD Lucia}.} \bibinfo{year}{2018}\natexlab{}.
\newblock \showarticletitle{Automatic test smell detection using information
  retrieval techniques}. In \bibinfo{booktitle}{\emph{Proceedings of the
  International Conference on Software Maintenance and Evolution (ICSME).
  IEEE}}.
\newblock


\bibitem[\protect\citeauthoryear{Spadini, Aniche, Bruntink, and
  Bacchelli}{Spadini et~al\mbox{.}}{2017}]%
        {spadini2017mock}
\bibfield{author}{\bibinfo{person}{Davide Spadini},
  \bibinfo{person}{Maur{\'\i}cio Aniche}, \bibinfo{person}{Magiel Bruntink},
  {and} \bibinfo{person}{Alberto Bacchelli}.} \bibinfo{year}{2017}\natexlab{}.
\newblock \showarticletitle{To Mock or Not To Mock? An Empirical Study on
  Mocking Practices}. In \bibinfo{booktitle}{\emph{Mining Software Repositories
  (MSR), 2017 IEEE/ACM 14th International Conference on}}. IEEE,
  \bibinfo{pages}{402--412}.
\newblock


\bibitem[\protect\citeauthoryear{Spadini, Aniche, Bruntink, and
  Bacchelli}{Spadini et~al\mbox{.}}{2019a}]%
        {spadini2019mock}
\bibfield{author}{\bibinfo{person}{Davide Spadini},
  \bibinfo{person}{Maur{\'\i}cio Aniche}, \bibinfo{person}{Magiel Bruntink},
  {and} \bibinfo{person}{Alberto Bacchelli}.} \bibinfo{year}{2019}\natexlab{a}.
\newblock \showarticletitle{Mock objects for testing java systems: Why and how
  developers use them, and how they evolve}.
\newblock \bibinfo{journal}{\emph{Empirical Software Engineering}}
  \bibinfo{volume}{24}, \bibinfo{number}{3} (\bibinfo{date}{Jun}
  \bibinfo{year}{2019}), \bibinfo{pages}{1461--1498}.
\newblock


\bibitem[\protect\citeauthoryear{Spadini, Palomba, Baum, Hanenberg, Bruntink,
  and Bacchelli}{Spadini et~al\mbox{.}}{2019b}]%
        {spadini2019test}
\bibfield{author}{\bibinfo{person}{Davide Spadini}, \bibinfo{person}{Fabio
  Palomba}, \bibinfo{person}{Tobias Baum}, \bibinfo{person}{Stefan Hanenberg},
  \bibinfo{person}{Magiel Bruntink}, {and} \bibinfo{person}{Alberto
  Bacchelli}.} \bibinfo{year}{2019}\natexlab{b}.
\newblock \showarticletitle{Test-driven code review: an empirical study}. In
  \bibinfo{booktitle}{\emph{Proceedings of the 41st International Conference on
  Software Engineering}}. IEEE Press, \bibinfo{pages}{1061--1072}.
\newblock


\bibitem[\protect\citeauthoryear{Spadini, Palomba, Zaidman, Bruntink, and
  Bacchelli}{Spadini et~al\mbox{.}}{2018}]%
        {spadini2018relation}
\bibfield{author}{\bibinfo{person}{Davide Spadini}, \bibinfo{person}{Fabio
  Palomba}, \bibinfo{person}{Andy Zaidman}, \bibinfo{person}{Magiel Bruntink},
  {and} \bibinfo{person}{Alberto Bacchelli}.} \bibinfo{year}{2018}\natexlab{}.
\newblock \showarticletitle{On the relation of test smells to software code
  quality}. In \bibinfo{booktitle}{\emph{Proceedings of the International
  Conference on Software Maintenance and Evolution (ICSME). IEEE}}.
\newblock


\bibitem[\protect\citeauthoryear{{van Deursen}, Moonen, Bergh, and Kok}{{van
  Deursen} et~al\mbox{.}}{2001}]%
        {Deursen01refactoringtest}
\bibfield{author}{\bibinfo{person}{Arie {van Deursen}}, \bibinfo{person}{Leon
  Moonen}, \bibinfo{person}{Alex Bergh}, {and} \bibinfo{person}{Gerard Kok}.}
  \bibinfo{year}{2001}\natexlab{}.
\newblock \showarticletitle{Refactoring Test Code}. In
  \bibinfo{booktitle}{\emph{Proceedings of the 2nd International Conference on
  Extreme Programming and Flexible Processes in Software Engineering (XP)}}.
  \bibinfo{pages}{92--95}.
\newblock


\bibitem[\protect\citeauthoryear{White and Marsh}{White and Marsh}{2006}]%
        {white2006content}
\bibfield{author}{\bibinfo{person}{Marilyn~Domas White} {and}
  \bibinfo{person}{Emily~E Marsh}.} \bibinfo{year}{2006}\natexlab{}.
\newblock \showarticletitle{Content analysis: A flexible methodology}.
\newblock \bibinfo{journal}{\emph{Library trends}} \bibinfo{volume}{55},
  \bibinfo{number}{1} (\bibinfo{year}{2006}), \bibinfo{pages}{22--45}.
\newblock


\bibitem[\protect\citeauthoryear{wiki}{wiki}{2019}]%
        {mozillaSheriff}
\bibfield{author}{\bibinfo{person}{Mozilla wiki}.}
  \bibinfo{year}{2019}\natexlab{}.
\newblock \bibinfo{title}{Sheriffing}.
\newblock \bibinfo{howpublished}{\url{https://wiki.mozilla.org/Sheriffing}}.
\newblock


\bibitem[\protect\citeauthoryear{Wohlin}{Wohlin}{2014}]%
        {wohlin2014guidelines}
\bibfield{author}{\bibinfo{person}{Claes Wohlin}.}
  \bibinfo{year}{2014}\natexlab{}.
\newblock \showarticletitle{Guidelines for snowballing in systematic literature
  studies and a replication in software engineering}. In
  \bibinfo{booktitle}{\emph{Proceedings of the 18th international conference on
  evaluation and assessment in software engineering}}. ACM,
  \bibinfo{pages}{38}.
\newblock


\bibitem[\protect\citeauthoryear{Zhang, Jalali, Wuttke, Muslu, Lam, Ernst, and
  Notkin}{Zhang et~al\mbox{.}}{2014}]%
        {Zhang:2014:ERT}
\bibfield{author}{\bibinfo{person}{Sai Zhang}, \bibinfo{person}{Darioush
  Jalali}, \bibinfo{person}{Jochen Wuttke}, \bibinfo{person}{Kivan{\c{c}}
  Muslu}, \bibinfo{person}{Wing Lam}, \bibinfo{person}{Michael~D. Ernst}, {and}
  \bibinfo{person}{David Notkin}.} \bibinfo{year}{2014}\natexlab{}.
\newblock \showarticletitle{Empirically Revisiting the Test Independence
  Assumption}. In \bibinfo{booktitle}{\emph{Proceedings of the International
  Symposium on Software Testing and Analysis (ISSTA)}}.
  \bibinfo{publisher}{ACM}, \bibinfo{pages}{385--396}.
\newblock
\showISBNx{978-1-4503-2645-2}
\urldef\tempurl%
\url{https://doi.org/10.1145/2610384.2610404}
\showDOI{\tempurl}


\end{thebibliography}

\end{document}